\documentclass[prl,showpacs,showkeys,floatfix]{revtex4}

\usepackage{bm}
\usepackage{graphics}
\usepackage{graphicx}
\usepackage[dvips,usenames]{color}

\newcommand{\vs}{\vspace{4mm}}

\newcommand{\be}{\begin{equation}}
\newcommand{\ee}{\end{equation}}
\newcommand{\ba}{\begin{eqnarray}}
\newcommand{\ea}{\end{eqnarray}}
\newcommand{\NL}{\nonumber \\}

\newcommand{\AuthorTeam}{
 \author{T.~S.~Bir\'o\footnote{Lecture for students at QM2005}} 
 \affiliation{
   Institute for Theoretical Physics, University of Giessen,
   D 35392 Heinrich-Buff-Ring 16, Giessen, Germany
  }
 \affiliation{
  KFKI Research Institute for Particle and Nuclear Physics,
  H 1526 Budapest Pf. 49, Hungary
 }
}

\begin{document}

\title{
Introduction to statistical models and non-extensive statistics
}

\AuthorTeam



\vs
\begin{abstract}
 Predictions for occurence of a quark matter in heavy ion collisions
 were made up to now in the framework of extensive thermodynamics.
 We review here some basic concepts in statistics, kinetic theory and
 thermodynamics, in particular models establishing a non-extensive
 statistics. Possible connections to particle spectra emerging from 
 high-energy reactions are reviewed.
\end{abstract}

\maketitle


\section{Introduction}

\vs
Statistical models in general apply to phenomena which appear in
a sufficiently large number and can be repeated independently and
(in principle) indefinitely. Whether high energy physics and in particular
heavy-ion collisions provide such phenomena, is a question to be discussed on
its own right. We shall try to contribute to an answer to this question
here by reviewing some basic concepts of
statistics, kinetic theory and thermodynamics, with the aim to clarify
the limits of the application of statistical models.

\vs
First we review the central limit theorem of statistics\cite{CENTRAL}, 
as the basic law governing the result of very many independent influences.
Then kinetic theory, designed to describe the way towards equilibrium
and its maintenance,
is discussed. By doing so we point out quite a few modern applications 
of statistical methods describing
a stationary state, which obeys  non-extensive thermodynamical rules\cite{NEXT-THERMO}.
A review of non-extensive rules, equilibrium one-particle distributions
and entropy density formulas follows.

\vs
An underlying particular application we have in mind is a relativistic
heavy ion collision\cite{RHIC}. The first touch physics is probably dominated by
nonlinear field dynamics and parton collisions. While some very energetic
partons may escape in form of jets, in a central heavy-ion event most of
the beam energy is transformed into a compression of the nuclear matter 
and production of
relativistic quarks and gluons. How a quark-gluon plasma in a thermal state
is formed in these events and with which properties,
is still an objective of the contemporary research.

\vs
Assuming i) such a quark matter formation in an intermediate
state, ii) ergodization of energy between many newly produced particles and
iii) a relatively fast hadronization, the experimentally measured specific hadron
spectra may reflect statistical, presumably even equilibrium thermal,
properties of the precursor matter\cite{STAT-MOD}. Strong final state interaction on the
other hand, a so called hadronic afterburner, 
may wash out spectral characteristics of earlier stages of the evolution. 
Whether this is the case, can in principle be
studied by checking quark coalescence rules or comparing hadron- and 
lepton-spectra\cite{ALCOR}.  
Finally the late resonance decay\cite{RESONANCES} during free
streaming of hadrons changes the hadrochemical composition. 
Fortunately, some properties (e.g.  the transverse
momentum spectra) are influenced only partially (at their low end) by this.
Since the relativistic energy is given by $E=m_T \cosh y$ with
transverse mass $m_T=\sqrt{p_T^2+m^2}$ and rapidity $y$ for a particle with mass
$m$, the best way to study statistical equilibrium distribution of
hadrons is the comparison of $m_T$-spectra at rapidity $y=0$ 
for different particles.
A universal behavior ($m_T$-scaling)\cite{MT-SCALING} indicates that the one-particle
distributions depend on the energy only and not on all momentum
components: a basic feature of generalized and conventional
thermal distributions.

\vs
In this lecture we review general basic ideas and mathematical formulas
related to statistical models. By doing so emphasis is given to particular
cases when the conventional picture of earlier textbook statistical physics
does not apply, instead some generalized concepts have to be considered.
Starting with the central limit theorem of statistics we continue with
a discussion of the Brownian motion in the framework of the Langevin and
Fokker-Planck equations. Here the very level of generality is identified
that leads to a Tsallis distribution instead of the Gibbs-Boltzmann 
one\cite{NEXT-MT-SCALING}.
The potential for further generalization is pointed out, too.
Then a discussion of the Boltzmann equation follows introducing to modern
generalizations capable to describe non-exponential 
equilibrium one-particle distributions\cite{NLBE,NEBE}. 
We close with some remarks about
how certain a thermodynamical state may be reconstructed from the observed
one-particle energy spectra.

\section{Statistics: the law of big numbers}
\vs

There is a mathematical property behind the applicability of
statistical physics: many "normal" distributions of probability
fold to a Gaussian with a width scaling down with increasing
number of individual constituents. This property is expressed
nicely in the central limit theorem\cite{CENTRAL}. An enlightening simple example
of its action is given by the distribution of the sum of uniform
random variables in a finite interval. Finally in this section an
example not falling under the reign of the central limit theorem,
the Lorentzian distribution, is presented\cite{ALMOST-EXP,LEVY}.


\vs
Let $x_i$ be random distributed according to $w_i(x_i)$.
We are interested in the distribution of a scaled sum of $n$ such
variables:
\be
 P_n(x) \: = \: \int  \prod_{i=1}^n w_i(x_i) \,
 \delta\left( x - a_n \sum_{k=1}^n x_k \right).
\ee
Here we assumed that the joint probability of having $n$ values
is a product of the individual probabilities; this is the requirement
of statistical independency. With this assumption the Fourier transform
of the seeked probability is a product of properly scaled Fourier transforms of
individual probabilities:
\be
 \tilde{P}_n(k) = \int\! dx \, e^{ikx} P_n(x) \: = \: \prod_{i=1}^n \tilde{w}_i(a_nk).
\ee
From the Taylor expansion of $\ln \tilde{P}(k)$ around $k=0$
one obtains the central moments (correlations). For the $\ell$-th
moment the scaling law, 
$\sigma_n^{(\ell)}=a_n^{\ell} n \overline{\sigma}^{(\ell)}$
applies with $\overline{\sigma}$ being the finite average of the given
central moment of the individual distributions. Whenever
$\sigma_i^{(\ell)}=0$ for all $\ell<\ell_0$ values and
$\sigma_i^{(\ell_0)}$ is finite, all the higher moments can be
scaled down in the folded distribution by choosing $a_n \propto n^{-1/\ell_0}$.
The resulted scaling,  
$\sigma_n^{(\ell)}= n^{1-\ell/\ell_0} \overline{\sigma}^{(\ell)}$,
in the $n \rightarrow \infty$ limit leaves us with only one
nonzero central moment, the $\ell_0$-th one.

\vs
Usually the $\ell=0$ central moment is zero due to the normalization
of the probability: $\ln \tilde{P}(0) = \ln 1 = 0$.
The first moment ($\ell=1$) can be made zero due to symmetry or by
a simple shift in the variables $x_i$. The second central moment is
then the first nonzero value. As a consequence the resulted
distribution of the scaled sum has a second moment with all higher moments
vanishing, therefore $\ln \tilde{P}(k)$ is quadratic in $k$, so
$\tilde{P}(k)$ and with that $P(x)$ is Gaussian.


\vs
A nice, simple example is given for $x_i$-s uniformly distributed
in the interval $(-1,+1)$. The distribution of
\be
 x = \sqrt{\frac{3}{n}} \sum_{i=1}^n x_i
\ee
approaches the Gaussian:
\be
 \lim_{n\to\infty}\tilde{P}_n(k) = 
 \lim_{n\to\infty}\left(\frac{\sin(k\sqrt{3/n})}{k\sqrt{3/n}} \right)^n 
  \: = \: \exp(-k^2/2).
\ee


\vs
A counterexample is given by the Lorentzian distribution, having a Fourier
transform $\tilde{w}_i(k)=\exp(-|k|)$. Now there is a non-Gaussian
limiting distribution with an altered scaling for 
\be
 x = \frac{1}{n} \sum_{i=1}^n x_i.
\ee
It is itself a Lorentzian:
\be
 \lim_{n\to\infty} \tilde{P}_n(k) = \lim_{n\to\infty} \left(e^{-|k|/n} \right)^n 
  =  \exp(-|k|).
\ee
This is a special case of the L\'evy distribution\cite{LEVY}.


\vs
\section{Kinetic theory}

\vs
In a brief review of concepts distilled from the kinetic theory approach
to thermodynamics we shall rely on the notion of "noise" heavily.
The sum of random influences (forces) is itself a random variable.
It shows relationship to the sum of random numbers; its distribution
under quite general circumstances can be considered as a Gaussian
distribution. The independency of individual influences is assumed
first of all in time: such a view deals with uncorrelated stochastic
changes in the parameters describing a physical system.

\vs
The simplest physical theory of such a system is that of the Brownian
motion\cite{BROWN}, considering a free, massive particle under the influence of
forces acting as an uncorrelated, Gaussian (white) noise. A 
description of the motion of such a particle is given by the classical
Langevin equation, or equivalently a statistics over possible such motions
is described by the Fokker-Planck equation. A balance between 
damping and accelerating forces leads to a stationary state, with
vivid microscopical dynamics, but macroscopically (on the average
over many particles) presenting a thermodynamical equilibrium state.
General statements about this balance are comprised in the
fluctuation-dissipation theorem. In the next section we review generalizations
of the Boltzmann equation, a somewhat more complicated 
kinetic theory. The effects of such generalizations on the 
the concept and definition of entropy and on equilibrium
distributions will be an issue of a further section.

\vs
\subsection{General Langevin problem}

\vs
Let us consider a simple, one degree of freedom motion. The change
of momentum $p$ in time is given by a force depending on this
momentum and on a noise variable $z$:
\be
 \dot{p} = F(p,z).
\ee
Following the method pioneered by Ornstein and Uhlenbeck\cite{UHLENBECK} a distribution
of many possible values of $p$ at a time $t$ is considered.
This $f(p,t)$ distribution governs average values of a smooth
but otherwise arbitrary test function $R(p)$. The same integral over
$p$ can be expressed at the time $t+dt$ assuming a statistical average
over the noise $z$ in the time interval passed since $t$:
\be
\int\! dp \, R(p) f(p,t+dt) \: = \:
\int\! dp \, \langle \, R(p+dt\: F(p,z)) \, \rangle f(p,t).
 \label{WangUh}
\ee
One assumes that the averaging of the force $F$ over the noise $z$
gives the result:
\ba
\langle \: F \: \rangle &=& - G(p), \NL
\langle \: F \, F \: \rangle - \langle \: F \: \rangle \, 
\langle \: F \: \rangle &=& 2 D(p) / dt.
\ea
The above scaling of the correlation with $dt$ follows from the Gaussian
nature of the noise $z$.
Expanding now the equation(\ref{WangUh}) up to terms linear in $dt$
one arrives at:
\be
 \int\! dp \, R(p) \, \frac{\partial f}{\partial t} (p,t) \: = \:
 \int\! dp \, \left[ - G(p) R'(p) + D(p)R''(p)\right] f(p,t).
\ee
After partial integration and considering arbitrary $R(p)$ one gets
the Fokker-Planck equation:
\be
 \frac{\partial f}{\partial t} \: = \:
 \frac{\partial}{\partial p} \left( G \, f\right) \: + \:
 \frac{\partial^2}{\partial p^2} \left( D \, f \right).
\label{FP}
\ee

\vs
\subsection{Particular Langevin problem}

\vs
The above Langevin and Fokker-Planck problem is still quite general.
Damping and diffusion coefficients depend on the momentum $p$
in a general way. Ergodization in phase space is achieved on the
other hand when constant energy surfaces are covered. In such
a situation the distribution $f$, as well as the coefficients
$G$ and $D$ (the latter related to the noise), depends on the
energy $E(p)$ only. Note, however, that they are not constant.

\vs
Such a particular Langevin equation is given by\cite{SLIDING-SLOPE}
\be
\dot{p} = z - G(E) \frac{\partial E}{\partial p}
\ee
containing an energy dependent damping proportional to the 
general velocity $v=\partial E/\partial p$, and a 
zero-average noise, $\langle z(t)\rangle = 0$, with
a correlation
\be
 \langle z(t) z(t') \rangle = 2 D(E) \delta(t-t').
\ee
The Fokker-Planck equation contains in this case the factors
$D(p)=D(E)$ and $G(p)=-G(E)\partial E/\partial p$.
Its stationary solution is given by
\be
f(p) = \frac{A}{D(E)} \, \exp \left(- \int \frac{G(E)}{D(E)}dE \right) \: = \:
A \, \exp \left( - \int \frac{dE}{{\cal T}(E)}\right).
\ee
This result is not readily the Boltzmann-Gibbs distribution,
$\exp(-E/T)$, only in the case of energy-independent damping and
noise coefficients. A general equilibrium is able to feature almost
any other distribution of the energy of a single degree of freedom
picked out of its environment. Instead of a constant temperature, $T$,
in the general case a sliding inverse logarithmic slope is
characteristic to such states. From $1/{\cal T}(E)=-d \ln f(E)/dE$ 
its relation to the damping and diffusion coefficients follows:
\be
 {\cal T}(E) = \frac{D(E)}{G(E)+D'(E)}.
 \label{SLIDE}
\ee
The low-energy limit of this expression, pretending as $G(E)$ and $D(E)$
were constant, leads to an experimentally feasible definition of the
Gibbs temperature, $T_{Gibbs}=D(0)/G(0)$. From the viewpoint of the
Brownian motion another temperature may be used, the Einstein
temperature $T_{Einstein}=\lim_{E\to\infty} D(E)/G(E)$. None of these
two approximations are, however, coincident with the sliding slope
of particle spectra given by eq.(\ref{SLIDE}).

\vs
An important and historically mostly considered particular case is
given by the constant slope distribution, the Boltzmann-Gibbs
distribution: $T(E)=T$. A modern, non-classical distribution,
the Tsallis distribution seems to be the next simplest, having a
linear inverse slope -- energy relation:
\be
 {\cal T}(E) = T/q + E (1-1/q).
\label{LIN_SLOPE}
\ee
The corresponding equilibrium distribution in this case turns to be
an exponential of a logarithm, which is a power-law:
\be
 f(p) \: = \: \frac{1}{Z} \left(1+(q-1)\frac{E}{T} \right)^{\frac{q}{q-1}}
 \label{TSALLIS_DIST}
\ee
It has the interesting property, that the parameter $T$ is the fixed point
of the sliding (linear) slope: ${\cal T}(T)=T$. This parameter shall be 
referred to as the Tsallis temperature.

\vs
\subsection{Fluctuation-dissipation theorem}

\vs
In a realistic system there are many microscopic degrees of freedom to
be considered. Denoting a point in the $6N$-dimensional phase space by
$p_i$, the Langevin equation can be implemented in the form
\be
 \dot{p}_i \: = \: (S_{ij}-G_{ij}) \nabla_j E \: + \: z_i.
\ee
The symplectic coefficient, $S_{ij}$, does not change the total energy
of the system, $E(p_i)$, it causes conservative motion inside a given
energy shell only. For the sake of energy distribution it is
therefore not interesting and shall be omitted in the followings.
The damping and dissipation terms 
with the respective symmetric coefficients $G_{ij}$ and $D_{ij}$
keep balance on the long term.
An ergodized equilibrium distribution can be
a single function of the energy $E$ only, therefore these coefficient
matrices have to be connected by a single function of energy, too.
This gives rise to a general fluctuation dissipation theorem:
\be
 D_{ij}(E) \: = \: {\cal T}(E) \, \left( G_{ij}(E) + D'_{\! ij}(E) \right).
 \label{FT}
\ee
It is highly nontrivial that two high-dimensional matrix functions
of the phase space coordinates $p_i$ would be
related by a single scalar function of energy only!
Recalling that ${\cal T}(E)$ is the inverse logarithmic slope of the
equilibrium distribution, the fluctuation dissipation relation can
be expressed using $f(E)$, too:
\be
  D_{ij}(E) \: = \: \frac{1}{f(E)} \: \int\limits_E^{\infty} \, 
  G_{ij}(x) \, f(x) \, dx.
\ee
Still, quite general diffusion and damping coefficient matrices are allowed,
but $D_{ij}$ is connected to $G_{ij}$ via an energy dependent scalar,
the equilibrium energy distribution, $f(E)$. Particular cases of this
relation are i) the use of the Gibbs distribution, $f(E)\propto\exp(-E/T)$,
leading to $D_{ij}=TG_{ij}$ with constant matrices (the usual textbook case),
or ii) the use of a Tsallis distribution (\ref{TSALLIS_DIST}) giving rise
to $D_{ij}(E)=\left( T +(q-1)E \right) G_{ij}$ with constant $G_{ij}$ but
linearly energy dependent diffusion coefficient, $D_{ij}$.
A further interesting case is presented by assuming that both the damping
and the diffusion coefficient matrix is a linear function of the energy,
but their ratio (the Einstein temperature) is constant. This assumption is
typical to field theory calculations. Due to eq.(\ref{FT}) the sliding slope
is, however, not constant; it rather interpolates between a linear rise at
low energy and a constant at high energy. From $D=TG=\gamma T(1+E/E_c)$ it follows
$ 1/{\cal T}(E) \: = \: 1/T + 1/(E+E_c)$.

\vs
\subsection{Additive and multiplicative noise}

\vs
Another approach to reach a non-exponential stationary energy distribution
from a kinetics described by the Langevin equation considers a stochastic
damping coefficient. The Langevin equation is kept linear,
\be
 \dot{p} = \zeta - \gamma p,
\ee
but, more general than in the classical approach, both $\zeta$ and $\gamma$
are stochastic variables\cite{MULT-NOISE}. 
With constant mean values, $\langle \zeta \rangle =F$
and $\langle \gamma \rangle = G$, and white-noise correlations,
\be
\langle \gamma(t) \gamma(t') \rangle = 2 C \, \delta(t-t'), \qquad
\langle \zeta(t) \gamma(t') \rangle = 2 B \, \delta(t-t'), \qquad
\langle \zeta(t) \zeta(t') \rangle = 2 D \, \delta(t-t'),
\ee
the equivalent Fokker-Planck equation (\ref{FP}) contains a damping factor 
$G(p)=Gp-F$ and a diffusion factor $D(p)=Cp^2-2Bp+D$. For $B\ne 0$ the
two noisy coefficients are cross-correlated. For a single degree of freedom
the stationary distribution can be obtained analytically:
\be
 f(p) = A \left(\frac{D}{D(p)} \right)^v
 \, \exp \left( - \frac{\alpha}{\theta} {\rm atan} (\frac{p\theta}{D-Bp} ) \right)
\ee
with the power $v=1+G/2C$, the exponent factor $\alpha=GB/C-F$ and the
variable $\theta=\sqrt{DC-B^2}$. For $F=0$ a characteristic momentum scale
in this distribution is given by $p_c^2=D/C$, the ratio of the additive and
multiplicative noise widths. For $F=0$ and $B=0$, i.e. for two independent
noises, the Tsallis distribution arises as a stationary solution:
\be
 f(p) = A \left( 1 + \frac{C}{D} p^2 \right)^{-v}.
\ee
Utilizing the energy formula for a free, massive, non-relativistic particle
it reads as
\be
 f(p) = A \left(1 + (q-1)\frac{E}{T} \right)^{-\frac{q}{q-1}}
\ee
with the Tsallis index $q=1+2C/G$ and the Tsallis temperature coinciding with
that of the classical Brownian motion, $T=D/mG$.
In the small momentum limit, $p\ll p_c$ the Tsallis distribution is nearly
Gaussian, independent from the properties of the multiplicative noise,
$f=A\exp(-Gp^2/2D)$ (or expressed with the energy a Gibbs distribution,
$f=A\exp(-E/T)$). In the opposite limit at high energy it is a power-law
distribution, $f=A(p/p_c)^{-2v}=A(E/E_c)^{-v}$. It is interesting to note,
that a definite relation arises between the energy scale, $E_c$, the temperature
parameter, $T$, and the tail power, $v$:
\be
 v = 1 + E_c/T.
 \label{TEST}
\ee
This relation can be experimentally tested.

\vs
\section{Non-Extensive Boltzmann Equation}

\vs
The heart of kinetic theory is the classical Boltzmann equation.
It does not only describe dynamical evolution of large systems
microscopically, resulting in stationary distributions on which 
thermodynamics can be based on, but it also offers a microscopical
foundation to the key quantity entropy. It has, of course, also
quite a few assumptions built in the theory; dropping one or other
of them may lead to a generalization of the classical approach.
The most recognized assumption, the micro-reversibility of the
transition probability, establishes H-theorem and the definition
of entropy; it should not be dropped. Less explicit assumptions,
like taking the two-particle probability as a product of one-particle
probabilities, or taking the total energy of a colliding pair as
the sum of the respective one-particle energies for freely moving
(asymptotic) particles, can be more readily generalized.

\vs
We review these two generalizations of the Boltzmann equation
in this section: the generalization of the product rule for
probabilities (dropping statistical independency) leads to a
non-linear Boltzmann equation (NLBE\cite{NLBE}), while considering
two-particle energies composed by an extended addition rule
mounds in the non-extensive Boltzmann equation (NEBE\cite{NEBE}).
Resulting stationary distributions, the H-theorem and
the main characteristics of generalized thermodynamicses
following from this will be presented.

\vs
\subsection{NLBE: generalized product}

\vs
The general structure of the Boltzmann equation describes the evolution
of the one-particle phase space density (interpreted as finding
probability of a particle or a microstate) via an integral over all
possible transitions to and from other states:
\be
 \dot{f}_1 \: = \: \int \limits_{234} \, w_{1234} \,
 \left( f_{34,12} - f_{12,34} \right).
 \label{NLBE}			  
\ee
Here the dot denotes a total time derivative (Vlasov operator) comprising
the essential evolution of the one-particle phase space density, $f_1$.
The indices ${1234}$ refer to two particles before and after a microcollision.
The transition probability, $w_{1234}$ reflects microreversibility and partner symmetry
in the permutation of these indices. It is positive and it contains
some conditions on conserving physical quantities; at least momentum and
energy.
\be
 w_{1234} = M^2_{1234} \, \delta((\vec{p}_1+\vec{p}_2)-(\vec{p}_3+\vec{p}_4)) \, 
                          \delta(E_{12}-E_{34}),
\ee
with $E_{12}$ total two-particle energy before and $E_{34}$ after the
collision.  The particle density factors,
$f_{12,34}$ and $f_{34,12}$ weight the transition yields for a 
$3+4\to 1+2$ and for a $1+2\to 3+4$ process, respectively.

\vs
In the NLBE approach the traditional simple additivity of energy is kept,
\be
 E_{12} = E_1 + E_2,
\ee
but the classical product formula of Boltzmann, $f_{12,34}=f_1f_2$, or the
supplement with blocking factors due to Uehling and Uhlenbeck
$f_{12,34}=f_1f_2(1\pm f_3)(1\pm f_4)$ is generalized. The generalization
still reflects particle separation property (particle $1$ goes into particle $3$
and particle $2$ goes into particle $4$ classically), but abandons the
linearity:
\be
 f_{12,34} = \gamma(f_1,f_3) \cdot \gamma(f_2,f_4).
\ee
This intends to simulate statistical correlations between initial and final
states and a nonlinearity of transition yields. Further requirement is that
the phase space density factor, $\gamma$ factorizes to a production
factor $a$, to a blocking factor $b$ (depending on the respective initial
and final phase space densities only) and to a factor symmetric in both:
\be
 \gamma(x,y) \: = \: a(x) \: b(y) \: c(x,y)
\ee
with $c(x,y) = c(y,x)$. 

\vs
The stationary state of the NLBE is governed by the ratio $\kappa(x)=a(x)/b(x)$.
It is easy to see from the following derivation of the generalized H-theorem.
We seek for a quantity called entropy in form of a one-particle phase space
integral, $S=\int\limits_1 \sigma(f_1)$. The question is what $\sigma(f)$
functional form guarantees macro-irreversibility, i.e. $\dot{S}\ge 0$.
Using the general Boltzmann equation (\ref{NLBE}) one writes
\be
 \dot{S}=\int\limits_1 \dot{f}_1 \sigma'(f_1) \: = \:
 \int\limits_{1234} w_{1234} \, c_{13} \, c_{24} \, \sigma'_1 \: 
 (a_3b_1a_4b_2-a_1b_3a_2b_4),
\ee
with the corresponding indices referring to arguments of the functions
$a$, $b$ and $c$. Now we explore particle permutation symmetries of this
expression. After the exchange of particle $1$ with $2$ and simultaneously
particle $3$ with $4$ we describe the same process. An exchange of the
initial with the final state is done by $1\leftrightarrow 3$ 
and $2\leftrightarrow 4$ (micro-reversibility). It amounts to a relative minus
sign in the total rate by exchanging gain and loss terms. Finally
the combined operation of both also contributes with a minus sign in total.
Dressing now the transition rate with factors carrying the same symmetry
we use $\tilde{w}_{1234}=w_{1234}c_{13}c_{24}b_1b_2b_3b_4/4$. With the notation
$\kappa_i=a_i/b_i$ for the production to blocking ratios we arrive at
\be
 \dot{S} = \int\limits_{1234} \tilde{w}_{1234}
 \left( \sigma_1'+\sigma_2' - \sigma_3' - \sigma_4'\right)
 \left( \kappa_3\kappa_4-\kappa_1\kappa_2\right).
\ee
This already reminds to the structure of the H-theorem result of the classical
Boltzmann equation. The correct entropy density can be read off as satisfying
\be
 \sigma'(f) = - \ln \kappa(f).
\ee
The stable equilibrium, where -- due to ergodization -- $f(p)$ can be expressed as a 
solely function of the corresponding energy, satisfies a product rule
for the $\kappa(f_i)$-s while the energy is additive. The only solution is that
the canonical equilibrium of the NLBE is given by
\be
 \kappa(f) = \frac{1}{Z} \exp(-E/T).
\ee
As particular cases the well-known Boltzmann-Gibbs, Fermi-Dirac or
Bose-Einstein distributions are recovered, but this result is far more
general. The corresponding general entropy formula can be constructed
founding a generalized thermodynamics with (in general) non-extensive
entropy composition rules when merging large subsystems.

\vs
\subsection{NEBE: generalized sum}

\vs
Another, recently pursued way is to keep the statistical independency, $f_{12,34}=f_1f_2$,
but to generalize the energy addition formula to a nontrivial composition rule
\be
 E_{12} = h(E_1,E_2).
\ee
Rules not being a simple sum, $h(x,y)\ne x+y$, present a non-extensive
energy composition. Latest in the thermodynamical limit, only
associative rules are physical: $h(h(x,y),z)=h(x,h(y,z))$. Due to a mathematical
theorem\cite{MATH} for functional equations the general solution of the associativity
requirement is a strict monotonous mapping to the simple addition:
\be
 X(h) = X(x) + X(y).
\ee
This is unique up to a constant factor. 

\vs
As a consequence for any microcollision $X(E_1)+X(E_2)=X(E_3)+X(E_4)$ holds.
The stationary solution of NEBE is hence given by
\be
 f(p) = \frac{1}{Z} \exp(-X(E)/T)
\ee
allowing again for a general, non-exponential energy dependence. It is
noteworthy that to each associative non-extensive composition rule $h(x,y)$ there
exist a mapping function $X(E)$. An in-medium dispersion relation, or
quasi-energy is obtained this way: the total sum,
\be
 X(E_{tot}) = \sum_i X(E_i)
\ee
is conserved by the NEBE.

\vs
In order to connect the approaches NLBE and NEBE (and both with the non-extensive
thermodynamics) the non-extensive composition rule of the energy has to be
extended to composition rules of other, traditionally "extensive" quantities.
Most important is the entropy composition rule. And, as putting a corner stone
into the right place, the scaling law
\be
 X(E)/T = X_s(E/T)
\ee
allows us to relate the equilibrium distributions by a one-variable
functional relation,
\be
 Zf_{eq}(E) = \exp(-X_s(E/T)) = \kappa^{-1}(\exp(-E/T)).
\ee
At the same time the generalized entropy density satisfies
\be
 \sigma'(f) = - \ln \kappa(Zf) = X_s^{-1}(-\ln Zf)
\ee
in equilibrium. This gives rise to an interpretation of the H-theorem for NEBE
where the never-decreasing total entropy is given by the same strict
monotonic back-mapping from the original Boltzmann-entropy:
\be
 S_B \: = \: X_s(S_{tot}) = \int -f\ln f.
\ee
The use of a non-extensive entropy composition rule,
$h_s(x,y)$, mapped to additivity by  $X_s(t)$ 
leads to the following mapping of the individual entropy-density: 
$s(f) = f X_s^{-1}(-\ln f)$.
The usual non-extensive entropy is then defined by $S_T=\int s(f)$.

\vs
\section{Thermodynamicses}

\vs
It is enlightening to list some particular cases of non-extensive composition laws
and corresponding thermodynamicses\cite{NEBE+NLBE}. 
The trivial law, $h(x,y)=x+y$, is mapped
by the identity, $X(E)=E$, and in (canonical) equilibrium the Boltzmann-Gibbs
distribution, $\exp(-E/T)$, emerges. The entropy density is 
given by $s(f) = -f \ln f$.

\vs
Tsallis' non-extensive thermodynamics relies on the composition rule,
$ h(x,y)=x+y+axy $,
connected to the mapping 
\be
 X_s(S) = \frac{1}{aT} \ln \left( 1+aT \, S\right),
\ee
first presented by Abe\cite{ABE}. The canonical equilibrium distribution,
\be
 f_{eq}(E) = \frac{1}{Z} \left( 1+aE\right)^{-1/aT},
\ee
can easily be connected to the Tsallis distribution (\ref{TSALLIS_DIST})
by $q=1-aT$.  Finally an entropy density
following Tsallis' original suggestion (encountered also earlier by several 
authors\cite{OTHERS}):
$ s(f) = \frac{f^q-f}{q-1}$,
can be obtained. 

\vs
An interesting endeavor is to consider the composition rule,
$ h(x,y) = \left( x^b + y^b \right)^{1/b}$.
Mapping the energy to its power, $X(E)=(aE)^b/a$, it gives rise in equilibrium
to a L\'evy distribution\cite{LEVY}, a.o. known from anomalous diffusion problems:
\be
 f_{eq}(E) = \frac{1}{Z} \exp\left(-(aE)^b/aT\right).
\ee
The corresponding expression for the entropy is an incomplete gamma function.

\vs
The pure multiplicative law, $h(x,y)=axy$, 
which may be considered as the high-energy limit of
Tsallis' construction with fixed energy scale $E_c=1/a$, 
is mapped by $X(E)=\frac{1}{a}\ln (aE)$ and leads to a pure power-law
behavior, $Zf_{eq}=(aE)^{-1/aT}$. 
The entropy formula due to $X_s(t)=(1/aT)\ln(aT \, t)$,
$ s(f) = \frac{1}{1-q} \, {f^q}$
reminds to the original suggestion by R\'enyi\cite{RENYI}.
In fact defining $S_R=X_s(S_T)$ the (Tsallis) entropy is mapped to the extensive
quantity:
\be
 S_R = \frac{1}{1-q} \: \ln \, \int f^q.
\ee

\vs
\subsection{Canonical and Extended Equilibrium}

\vs
Summarizing this part it seems to be useful to spell out the general conditions
leading to the traditional and to the generalized treatment of canonical
equilibrium. Three conditions are important:
\begin{enumerate}
\item	In the equilibrium state the energy shell is covered uniformly
	(ergodic assumption),
\item	The average energy exchange between the system under study and its
	environment is zero (not a driven system),
\item	The fluctuations in the system's energy are negligible compared to
	the average value of its energy (canonical limit).
\end{enumerate}
The third requirement has its physical basis in the fact that the characteristic
range of forces acting in the system is much less then the system size.
So in the thermodynamical limit of large systems, $V \gg A \ell$,
with a finite interaction range, $\ell$, all surface effects (proportional to $A$) 
in a large volume, $V$, become negligible. In this case the entropy is 
quite generally additive, $S_{12}=h(S_1,S_2)=S_1+S_2$, and the equilibrium
canonical distribution is of Boltzmann-Gibbs type, $Zf=\exp(-E/T)$.
Note that already the quantum statistical distributions violate this
requirement: the exchange interaction (Pauli blocking or Bose enhancement)
is long-ranged.

\vs
In an extended equilibrium state, non-extensive thermodynamics and generalized
kinetic descriptions deal with, the third requirement is not fulfilled.
Either long range forces, or large fluctuations (related to each other)
keep the non-extensivity parameter $a \sim A\ell/V$ finite. Applying this idea
to the time direction, also long-term memory effects spoil the traditional
picture of canonical equilibrium and textbook thermodynamics.
The unusual large fluctuations are treated in the framework of 
superstatistics\cite{SUPER-STAT},
where the traditionally intensive thermodynamical parameters, like temperature,
have a probability distribution instead of a fixed value. This view, of course,
also transforms the Boltzmann-Gibbs distribution into an arbitrary one:
\be
 f_{eq}(E) = \int\! d\beta \: {\cal P}(\beta) \: \frac{1}{Z(\beta)} \: e^{-\beta E}.
\ee
For example, the Tsallis distribution can be obtained by considering a Gamma-distributed
inverse temperature, $\beta=1/T$.
Some preview about a numerical simulation of NEBE for the Tsallis
distribution can be obtained from a poster presented by G.~Purcsel
at this conference.


\vs
\section{Hadron Statistics}

\vs
Statistical models have been often applied to hadron physics.
Starting with Rolf Hagedorn's statistical model of meson resonances\cite{HAGEDORN},
several attempts occured to describe hadron multiplicities in
elementary collisions by means of statistical distributions.
The very idea of a phase transition between confined and deconfined
quark matter relies on traditional equilibrium thermodynamics.
With the dawn of relativistic heavy ion physics a search for the
nuclear and quark matter equation of state began assuming a local
thermal equilibrium in an otherwise exploding fireball.
Such models and their predictions for experimentally observable
properties will be reviewed in other lectures\cite{TOR-LECT,CSER-LECT}.

\vs
\subsection{Particle spectra and equation of state}

\vs
Here, following the above introduction to non-extensive thermodynamics,
just some basic ideas about the interpretation of the 
power-law tails of nearly exponential spectra shall be discussed.
Both exponential and power-law regions are observed in hadron spectra
stemming from pp or heavy-ion collisions. The $m_T$-scaling property,
which reveals in particular after taking into account Doppler blue shift 
like corrections due to a transverse flow, indicates a thermodynamical
origin. Although the traditional approach explains the power-law tail
at very high $p_T$ values by pQCD calculations\cite{pQCD}, the non-extensive
statistics provides a unified view for the whole spectrum.
Certainly, jets also contribute to the hard part of light hadron spectra,
but their angular distribution is far from uniform. In principle,
these contributions may be filtered out. The question is, whether the rest
still shows a (now statistical) power-law behavior.

\vs
Another point can be made by inspecting the minimum bias pion $p_T$-spectrum
form RHIC AuAu collisions at 200 GeV (Fig.1 in Ref. \cite{SLIDING-SLOPE}).
To this a non-exponential fit can already be made at the $p_T$-region 
between $1$ and $4$ GeV. The extrapolation of this fit almost coincides
with the fit to the whole observed range between $1$ and $12$ GeV.
So one concludes that the power-law behavior is not only a very hard scale
physics. Furthermore in the non-extensive statistical approach there
is a connection between the soft properties (temperature $T$) and the
hard properties (power $v$, transition scale $E_c$) cf. eq.(\ref{TEST}). 

\vs
The dilemma between the statistical and pQCD explanation can be
well expressed in the following formula for the particle yield:
\be
 \frac{(2\pi\hbar)^3}{V} \, \frac{d^3N}{dk^3} \: = \:
 \int\limits_0^{\infty} \!d\omega \, \omega \, \rho(\omega,\vec{k}) \, f(\omega/T).
\ee
The observed spectra, even not considering a collective flow or an
extended source, are already convolutions of a spectral function,
$\rho(\omega,\vec{k})$, and statistical-enviromental effects, $f(\omega/T)$. 
Both these factors may embrance non-trivial effects, distorting this way a naively
expected Boltzmann-Gibbs exponential.
Both may contain a further energy scale; the spectral function
$\Lambda_{QCD}$, the thermodynamical weight $E_c$.
Even in the case, when $\rho$ would contain a very sharp peak at a given
dispersion relation, $\omega_k=\omega(\vec{k})$, only, the result is a composite function,
$\exp(-\omega_k/T)$, undistinguishable from a deformed thermodynamical
equilibrium, $\exp(-X(|k|)/T)$. Quite generally in this quasiparticle picture,
whenever the effective dispersion relation, $\omega(\vec{k})$, contains
parameters depending e.g. on the temperature, $T$, an assumption of an
exponential equilibrium distribution like $\exp(-\omega(\vec{k},T)/T)$ questions
the physical meaning of the temperature.
If it is an enviromental parameter, the inverse slope is not necessarily equal to this,
if it is an inverse slope, it is not necessarily the enviromental parameter
governing an effective dispersion relation.

\vs
Finally we note, that non-exponential distributions can also be interpreted
in the framework of so called superstatistics\cite{SUPER-STAT}. 
Here a distribution of the
intensive thermodynamical parameters is assumed instead of a fixed value.
The physical reason behind maybe the finiteness of the observed subsystem,
as well as extraordinarily large fluctuations not scaling down  
with the system size alike the ones governed by the Gaussian distribution would do.
The Tsallis distribution, for example, can be given by a continous distribution
of the inverse temperature according to a Gamma distribution:
\be
 (1+x/c)^{-(c+1)} \: =  \: \frac{1}{\Gamma(c+1)} \, \int\limits_0^{\infty} \!dt \,
 t^c \, e^{-t} \, e^{-xt/c}.
\ee
There are several possibilities to intrepret such a distribution.
Among others a heat conduction equation with multiplicative noise\cite{HEAT-COND}
or taking into account an energy imbalance in two-body collisions
due to a presence of further participant agents\cite{FLUCT-ENER}
leads to the required result.

\vs
\subsection{Limiting temperature with Tsallis distribution}

\vs
Another probe for the Tsallis distribution than the one-particle $p_T$-spectra
may be given by considering the spectrum of heavy hadronic resonances.
An exponentially growing mass spectrum, originally proposed by Hagedorn
and recently checked again latest experimental data in Ref.\cite{RES-SPECT},
with its famous consequence of having a limiting (or Hagedorn-) temperature
for such a system, can be reconstructed on the basis of Tsallis distributed
quark constituents. This approach\cite{ANDRE} assumes that the Tsallis distribution of
the quarks and antiquarks is folded into mesonic and baryonic distributions
of the conserved total energy satisfying $X(E)=\sum_i X(E_i)$.
In general for an $N$-fold convolution of massless constituents
with $d$ dimensional momenta one easily obtains that the average energy satisfies
\be
 \langle \, X(E) \, \rangle \: = \: N \, \sum_{j=1}^d \frac{TE_c}{E_c-jT}.
\ee
This expression diverges first as the temperature reaches the limiting value,
$T_H=E_c/d$ starting from zero (for positive $E_c=1/a$, i.e. for repulsive
modifications of the extensive energy composition rule). This way Hagedorn
hadrons emerge from Tsallis partons. This model also explains naturally, why the
baryonic and mesonic mass spectrum seems to have a different rise: they
contain different polynomial coefficients in front of three exponential factors.

\vs
\section{Conclusion}

\vs
In conclusion we have reviewed basic ideas of statistical physics and
kinetic theory which may lead to a non-extensive thermodynamics.
In particular the role of a multiplicative noise in the linear
Fokker-Planck and Langevin approach has been emphasized.
Generalizations of the Boltzmann equation towards non-factorizing
yield factors or non-additive energy composition rules were then
shown to lead again to non-extensive entropy definitions and
thermodynamics. Finally these ideas have been related to hadron
spectra observed in relativistic heavy ion collisions. We pointed out
that not only the omnipresence and $m_T$-scaling of power-law tails
in particle spectra, but also a possible interpretation of the
Hagedorn spectrum of heavy resonances may support the presence of
a non-exponential equilibrium distribution in hot quark matter.

\vs


\vs
{\bf Acknowledgment}

Enlightening discussions and common work 
with Drs. G\'eza Gy\"orgyi at E\"otv\"os University
and Antal Jakov\'ac at the Technical University Budapest are hereby
gratefully acknowledged. Further collaborations with Berndt M\"uller,
Andr\'e Peshier and G\'abor Purcsel contributed essentially to the
development of ideas presented above.
This work has been supported by the Hungarian National Science Fund OTKA
(T037689) and the Deutsche Forschungsgemeinschaft.



%


\end{document}